\documentclass[journal,draftclsnofoot,onecolumn,12pt]{IEEEtran}
\usepackage{amsthm,amssymb,graphicx,multirow,amsmath,color,amsfonts}
\usepackage[update,prepend]{epstopdf}
\usepackage[noadjust]{cite}
\usepackage[latin1]{inputenc}
\usepackage{tikz}
\usepackage{bbm} 
\usepackage{pdfpages}
\usepackage{multirow}
\usepackage{comment}

\def\nb0{{\mathbf{0}}}
\def\nb1{{\mathbf{1}}}

\allowdisplaybreaks 

\usepackage{setspace}	

\begin{document}
\pagenumbering{gobble}
\graphicspath{{./Figures/}}
\title{
Aerial Base Stations Deployment in 6G Cellular Networks using Tethered Drones: The Mobility and Endurance Trade-off
}
\author{
Mustafa A. Kishk, {\em Member, IEEE}, Ahmed Bader, {\em Senior Member, IEEE}, and Mohamed-Slim Alouini, {\em Fellow, IEEE}
\thanks{Mustafa A. Kishk and Mohamed-Slim Alouini are with King Abdullah University of Science and Technology (KAUST), Thuwal 23955-6900, Saudi Arabia (e-mail: mustafa.kishk@kaust.edu.sa; slim.alouini@kaust.edu.sa).\newline Ahmed Bader is with Insyab Wireless Limited, Dubai 1961, United Arab Emirates (e-mail: ahmed@insyab.com). This paper was accepted for publication in IEEE Vehicular Technology Magazine on Aug. 13th, 2020.
} 

}

\maketitle

\begin{abstract}
Airborne base stations (carried by drones) have a great potential to enhance coverage and capacity of 6G cellular networks. 
However, one of the main challenges facing the deployment of airborne BSs is the limited available energy at the drone, which limits the flight time. In fact, most of the currently used unmanned aerial vehicles (UAVs) can only operate for one hour maximum. This limits the performance of the UAV-enabled cellular network due to the need to frequently visit the ground station to recharge, leaving the UAV's coverage area temporarily out of service. In this article, we propose a UAV-enabled cellular network setup based on tethered UAVs (tUAVs). In the proposed setup, the tUAV is connected to a ground station (GS) through a tether, which provides the tUAV with both energy and data. This enables a flight that can stay for days. We describe in detail the components of the proposed system. Furthermore, we enlist the main advantages of a tUAV-enabled cellular network compared to typical untethered UAVs (uUAVs). Next, we discuss the potential applications and use cases for tUAVs. We also provide Monte-Carlo simulations to compare the performance of tUAVs and uUAVs in terms of coverage probability. For instance, for a uUAV that is available $70\%$ of the time (while unavailable charging or changing battery for $30\%$ of the time), the simulation results show that tUAVs with 120 m tether length can provide upto $30\%$ increase in the coverage probability, compared to uUAVs. Finally, we discuss the challenges, design considerations, and future research directions to realize the proposed setup.
\end{abstract}
\section*{Introduction} \label{sec:intro}
Drone-carried base stations (BSs) are believed to be an integral part of the 6G cellular architecture~\cite{8869705,dang_6g}. The inherent relocation flexibility and relative ease of deployment can be beneficial for multiple requirements of the next generation cellular networks, such as providing coverage in hotspots and in areas with scarce infrastructure such as disaster-recovering environments or rural areas. The higher probability to establish a line-of-sight (LoS) link with the ground users, because of the high altitude, leads to more reliable communication links and  wider coverage areas~\cite{7470933,7744808}. Potential use cases for airborne BSs include (i) offloading Macro BSs (MBSs) in urban and dense urban areas, and (ii) providing coverage for rural areas, which typically suffer from low cellular coverage due to lack of incentives for operators.

These potential advantages of airborne BSs have motivated the research community to study multiple aspects of UAV-enabled cellular networks such as the air-to-ground (A2G) channel characteristics, optimal placement of UAVs, and trajectory optimization~\cite{8641423}. In addition, there are two key design challenges in UAV-enabled systems that will be discussed in more details in this article. The first one is the limited energy resources available on board, which makes the flight time limited to less than one hour in most of the commercially available UAVs~\cite{8648453,8255733}. The second key design challenge is the wireless backhaul link~\cite{8255764}. 

Typically, the energy consumption of a UAV is two-fold: (i) propulsion energy, which is the energy consumed by the UAV for the purpose of flying and hovering and (ii) payload energy, which captures the energy consumption for communication and on-board processing. Many research works has been directed to designing energy efficient communication schemes for the UAVs in order to prolong their lifetimes. However, since propulsion energy is significantly more than the payload energy, energy efficient communication will not highly affect the flight time. Such short flight times might not be an issue for some use cases, such as drone-based delivery between nearby locations or data dissemination and collection from sensor networks. However, when it comes to establishing a UAV-mounted BS, longer flight times are vital in order to ensure a stable and uninterrupted cellular service. 


Unlike terrestrial BSs, which have wired backhaul links (typically using fiber cables), uUAVs rely on {\rm wireless backhaul} links. Compared to wired links, wireless backhaul links are susceptible to higher latency, interference, and lower achievable data rates. Hence, it is important to find the best technology to establish a wireless backhaul link at the uUAV~\cite{8255764}. Available solutions in the literature include: (i) satellite communication, (ii)  millimeter wave (mmWave) communication, (iii) free space optical (FSO) communication, and (iv) in-band backhaul communication. Each of these four solutions has its own pros and cons. For instance, satellite communication ensures a more reliable backhaul link, but suffers from higher latency. On the other hand, mmWave and FSO backhaul links ensure much higher data rates, compared to in-band backhaul. However, both solutions suffer from high vulnerability to blockage and are reliable only over small distances. The solution of using in-band backhaul receives most of the attention in current literature. This solution has lower latency, compared to satellite backhaul. It does not require an LoS channel to communicate efficiently, like mmWave or FSO. However, due to the high altitude of the uUAV, it suffers from higher levels of interference, which can reduce the achievable rate of the backhaul link significantly.
\begin{figure}
\centering
\includegraphics[width=0.6\columnwidth]{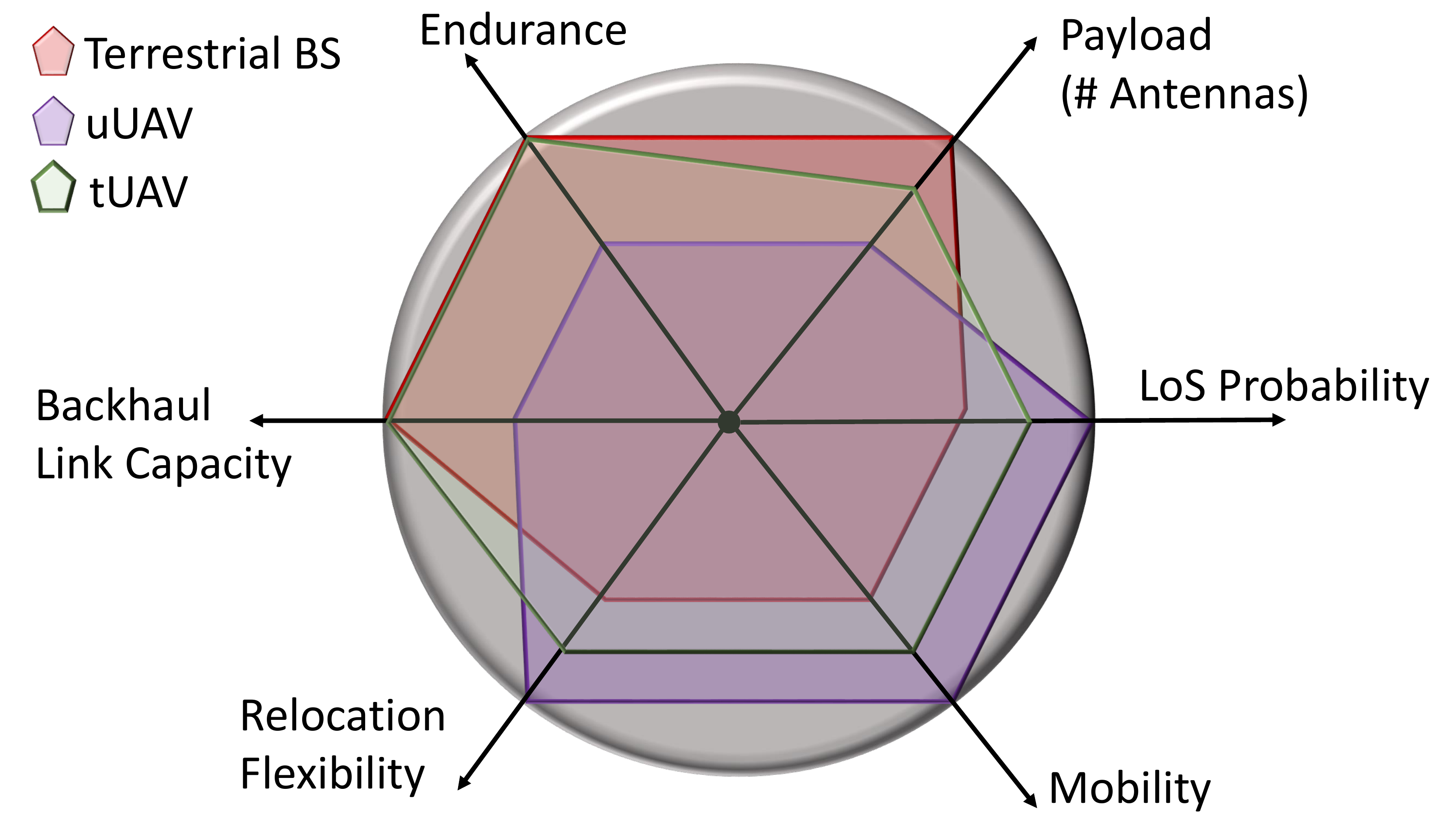}
\caption{A comparison between terrestrial BSs, uUAVs, and tUAVs.}
\label{fig:bridge}
\end{figure}

In this article, we propose a system setup that is based on tethered UAVs (tUAVs). The proposed technology solves the two technical challenges discussed above: (i) short flight time due to limited on-board energy, and (ii) establishing a reliable backhaul link. The interface between the GS and the tUAV is two fold: (i) energy supply and (ii) data link. The energy supply is provided from the GS to the tUAV through a wired connection, which enables the tUAV to sustain much longer flight times. Similarly, the data link between the GS and the tUAV is also physical through a fiber link, which ensures reliable communication at high data rates between the tUAV and the GS. Both the wired connections for energy and data are aggregated inside the tether. Currently, commercially available tUAVs can stay in the air with uninterrupted operation for days, with proven capability to tolerate harsh weather conditions. Due to its weight, the tether length is typically limited and ranges between 80 m and 150 m~\cite{kishk2019}. A recent incident in Puerto Rico have seen the deployment of a tUAV to provide cellular coverage for the suffered regions after Hurricane Maria~\cite{att}.

The main drawback in tUAVs is the limited tether length, which restricts the mobility and relocation flexibility of the drone. Hence, a trade-off naturally comes to picture between uUAV and tUAV as follows. On one hand, the tUAV has much longer flight time compared to uUAV due to the stable power supply through the tether. However, it can only hover or relocate within a restricted space defined by the tether length and the GS's surroundings. On the other hand, the uUAV has a complete freedom to hover and relocate anywhere in order to maximize the network performance. However, due to the limited on-board battery, it has to interrupt its operation regularly in order to recharge or replace its battery. Unfortunately, we do not have today the technology that can ensure long flight times while maintaining free mobility (tether-less). 
In Fig.~\ref{fig:bridge}, we visualize the key differences between terrestrial BSs, uUAVs, and tUAVs, in terms of main system advantages. It can be observed that the tUAV sacrifices the mobility and relocation flexibility of uUAVs in order to maintain the main requirements of a reliable cellular BS in terms of endurance and backhaul link quality. In the rest of this article, we will describe in detail the tUAV system. Next, we will discuss the main advantages of tUAVs and their potential applications and use cases. Finally, we will enlist the main challenges and design considerations that need to be carefully studied in future research works.
\begin{figure}
\centering
\includegraphics[width=0.7\columnwidth]{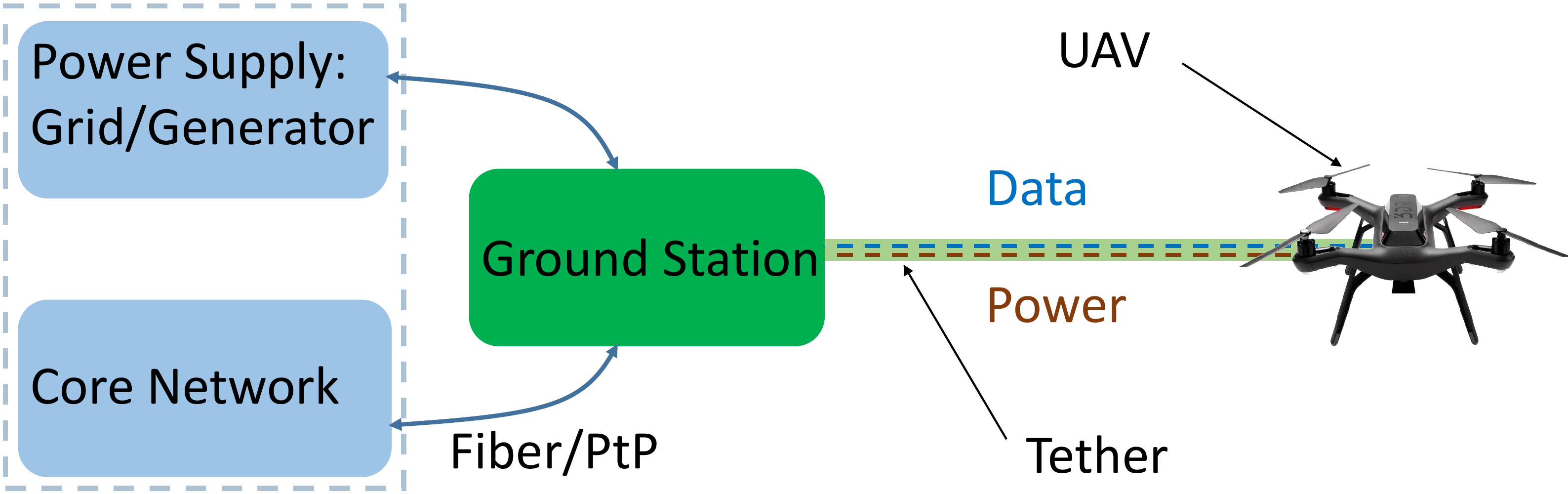}
\caption{Block Diagram of the considered tethered UAV system setup.}
\label{fig:setup}
\end{figure}
\section*{System Setup}
As shown in Fig.~\ref{fig:setup}, the proposed system setup consists of three main components: 
\begin{itemize}
\item the tUAV,
\item the tether,
\item the GS.
\end{itemize}
\noindent The GS is placed at a carefully selected location that satisfies two conditions: (i) has a reliable connection to the core network, and (ii) has a stable resource of energy such as the grid or a generator. These two connections (energy resource and core network) are extended to the tUAV through the tether. Hence, the tether provides the tUAV with uninterrupted energy supply, enabling it to stay in operation with significantly extended flight times. In addition, the tether also connects the tUAV to the core network through a wired connection providing it with a stable, reliable, and secure backhaul link. 
\begin{figure*}
\centering
\includegraphics[width=  7.4in, trim={1.2cm 0.4cm 1cm 0.78cm},clip]{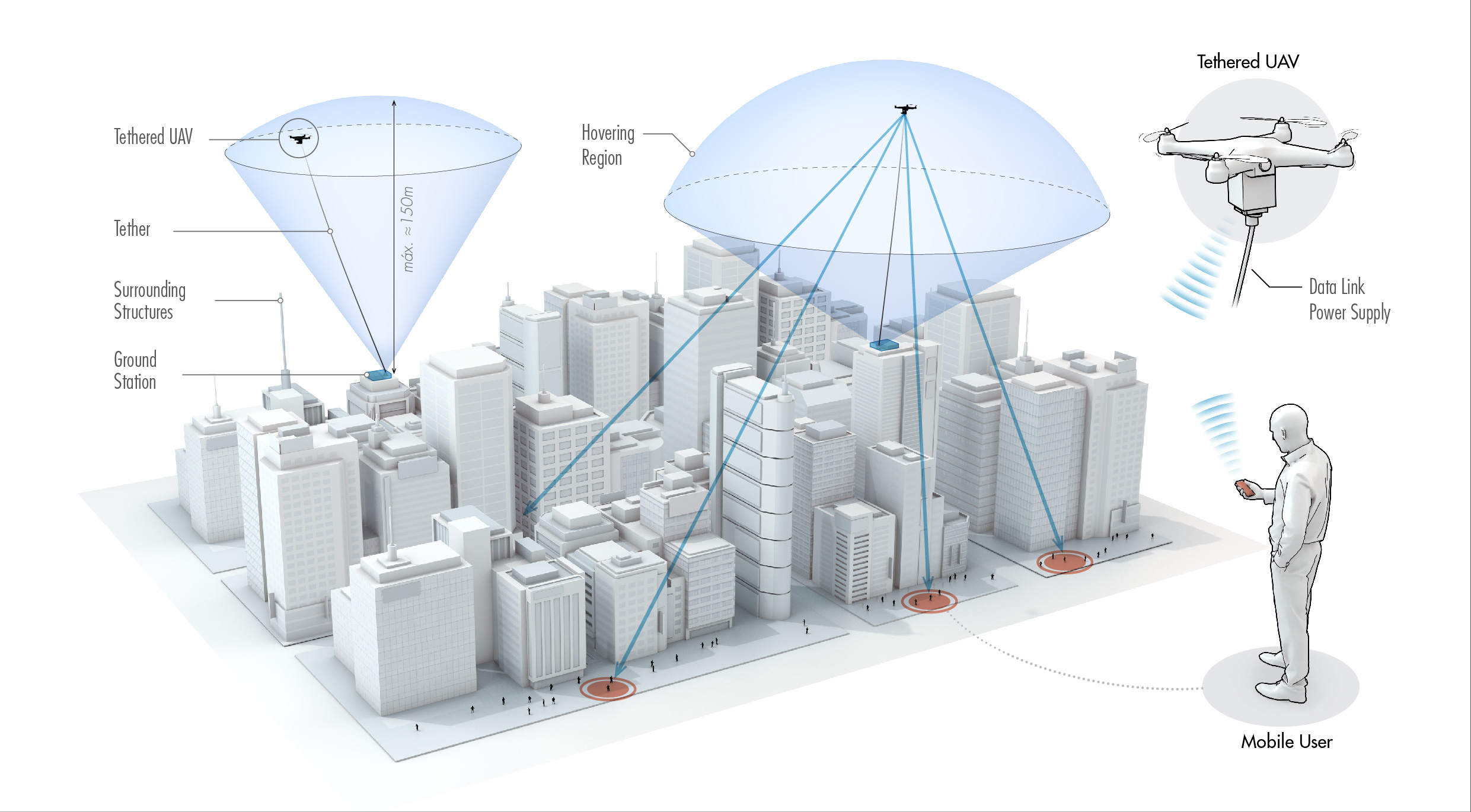}
\caption{tUAVs in urban areas.}
\label{fig:urban}
\end{figure*}

The tUAV can only hover within a specific range, which mainly depends on the tether length. Assuming the GS, which is the {\em launching point} of the tUAV, is placed at a rooftop, the tUAV can hover around the rooftop within a {\em truncated hemisphere} of radius equivalent to the tether length and centered at the rooftop, as shown in Fig.~\ref{fig:urban}. The overall region within which the tUAV can hover is limited by the heights of the neighboring buildings. Motion planning techniques can be adopted to determine the reachable 3-dimensional (3D) locations for a given environment, as discussed in~\cite{8594461}. In the rest of this article, we will refer to this region as {\em the hovering region}. 

The tUAV carries the antennas and a set of processing units. These processing units are connected to the GS through a data-carrying optical fiber along the tether. While the antennas and the processing units are considered heavy components for typical UAVs, current commercially available tUAV systems are able to carry up to 60 Kgs of additional payload~\cite{equinox}. The tUAV should hover within the hovering region and find the optimal 3D location that maximizes the cellular coverage for ground users. 

Beside its main job of providing the connection to the core network and the energy resource, the GS is responsible for controlling the tether. In particular, the GS should control the tension of the tether and ensure that it is taut at all times. During the tUAV's motion, the GS should sense if the tUAV requires releasing more length in order to reach its intended destination, or retracting extra length to ensure a taut tether~\cite{nicotra2014taut,7158825}.

It is clear from the above discussion that the smart selection of the GS's locations is of high importance for the performance of the tUAV system. For instance, placing the GS at a rooftop surrounded by taller buildings from all sides would reduce its hovering region to almost only the area above its own rooftop. Smaller hovering region leads to a more constrained tUAV 3D-placement problem and limits the mobility of the tUAV. The GS location selection process should take multiple aspects into consideration such as the traffic demand spatial distribution and the availability of the required infrastructure.

Aside from performance, the design of tUAV's system should take cost efficiency into consideration. There exists some differences, in terms of capital expenditure (CAPEX) and operational expenditure (OPEX), between tUAV and uUAV. CAPEX that exists in tUAV systems only, results mainly from (i) the tether and its mechanical controller, and (ii) the ground station. Meanwhile, CAPEX that exists in uUAV only results mainly from the charging stations required to recharge/replace the batteries of the uUAV. On the other hand, OPEX that only exists in tUAV systems mainly results from the rental of the rooftops that are used to deploy the ground station.

\section*{Main Advantages}
 \begin{figure*}
\centering
\includegraphics[width=  7.4in, trim={1cm 1.1cm 1cm 0.78cm},clip]{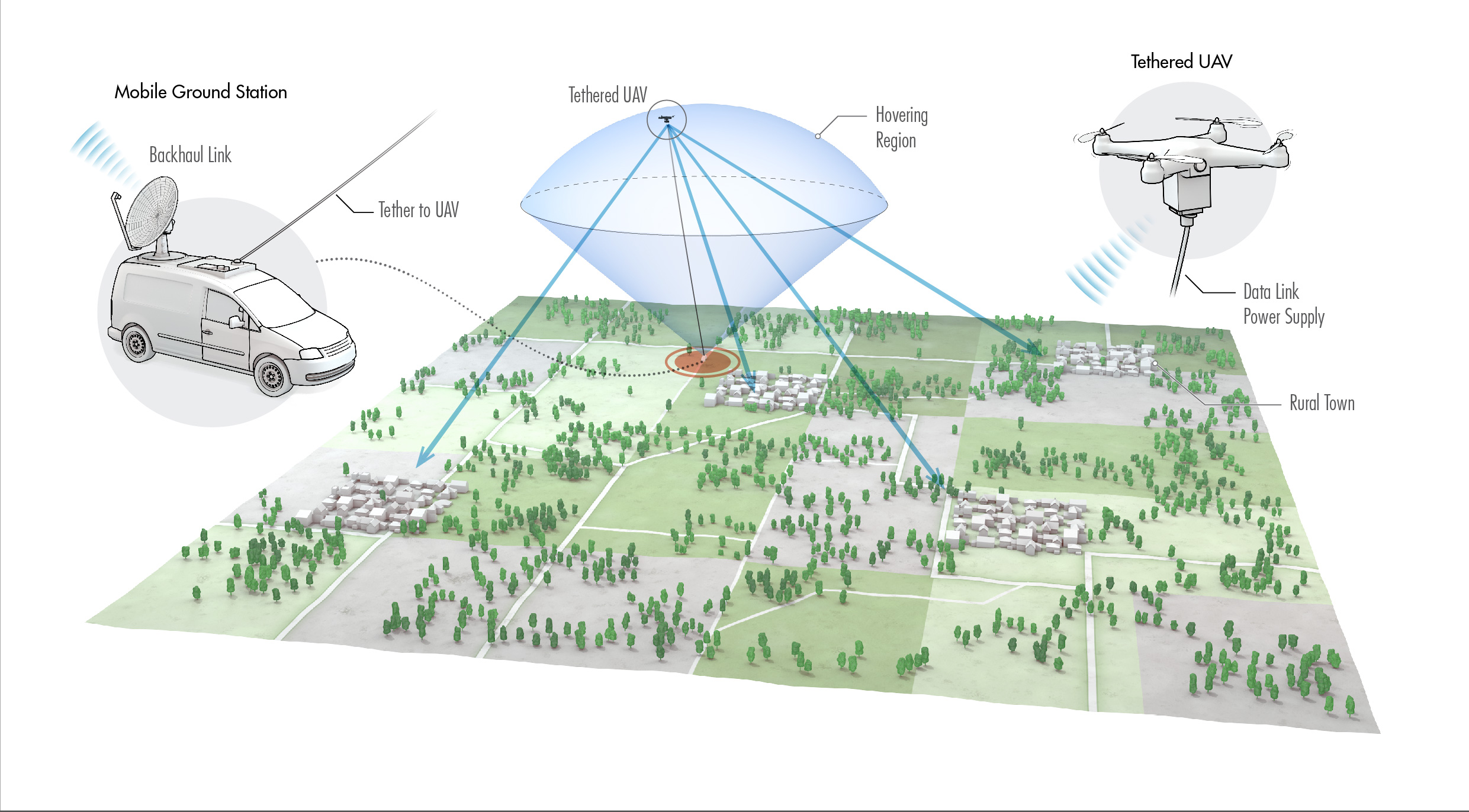}
\caption{tUAVs in rural areas.}
\label{fig:rural}
\end{figure*}
As discussed earlier, uUAVs are limited by the on-board battery as the sole resource of energy. Given that it typically takes less than an hour for battery depletion, the uUAV operation itself is quite limited in many aspects. For instance, the payload of the uUAV is typically kept low to reduce energy consumption during operation, the power dedicated for communication with ground stations or ground users is limited, and the relocation of the uUAV should be minimized since it consumes most of the available energy on-board. Hence, uUAV's energy limitations significantly affect its performance and reliability as a stable aerial BS. In this section, we will discuss in more detail the advantages of tUAVs compared to uUAVs or terrestrial BSs. 

\textbf{Advantage 1}. The tUAV can stay in operation for days. It needs to land at the GS only for maintenance, which is a normal procedure even for terrestrial BSs. Prolonged flight times of the tUAVs makes them comparable to terrestrial BSs in terms of endurance. However, tUAV has the advantage of higher altitude and more mobility (within the hovering range), which can be exploited to optimally place and relocate the tUAV according to the traffic demands and the channel conditions with mobile users. Furthermore, compared to terrestrial BSs, a tUAV has much safer maintenance procedure, since it does not require climbing high towers. One more advantage for the tUAV over terrestrial BS is the reduced terrestrial footprint. The space required for the GS can be as small as the rooftop of a typical urban building. This space is only required to place some processing units and to establish the connections to the energy resource and the core network. In addition, unlike the deployment of terrestrial BSs, which is fixed and permanent, the GS of the tUAV (which is its launching point) can be relocated whenever necessary from one rooftop to another. 

\textbf{Advantage 2}. The tUAV has the ability to sustain heavier payloads compared to uUAVs. This is due to the existence of a stable energy resource connected to the tUAV through the tether. In fact, current commercially available tUAV products can achieve up to 60 Kgs of payload~\cite{equinox}. 
Thus, tUAVs can afford having more antennas and radio chains, which enables sectorization and/or multiple-antenna communications. Hence, tUAVs offer more capacity and better interference management compared to uUAVs. 

\textbf{Advantage 3}. Due to having a wired data-link with the GS through the tether, tUAV can delegate some of the processing units to the GS, such as the baseband unit. This reduces the used tUAV payload, which enables placing more antennas at the tUAV.

\textbf{Advantage 4}. As explained earlier, one of the main research challenges in establishing uUAV as an aerial BS is backhaul communication. In fact, there is a trade-off between placing the uUAV close to the terrestrial BS to ensure strong and reliable backhaul link with high capacity, and placing the uUAV close to the mobile users to enhance the quality of the channel between the uUAV and mobile users. Hence, even in uUAV, free mobility and flexible placement is still constrained by the quality of the backhaul link and the distance to the terrestrial BS, which can be considered as {\em virtual tether}. On the other hand, tUAV has a stable wired connection to the GS with significantly higher capacity compared to a wireless backhaul link. Not only does this affect the achievable data rates, due to the high backhaul capacity, but also it frees more resource blocks for serving mobile users, that were reserved for wireless backhaul link in uUAV systems.

\textbf{Advantage 5}. One of the major concerns when using a uUAV is the {\em drone flyaway}. During uUAV operation, flyaway can be caused by many reasons such as software glitches, lost connection between the GS and the uUAV due to flying out of control range, hardware failure, interference in the communication channel leading to loss of control, or strong wind. In addition to the financial loss and the negative effect on the performance of the uUAV-enabled communication system, a drone flyaway imposes numerous public risks. It may crash into a pedestrian, a building, a highway, which might cause dangerous accidents. Many accidents caused by drone crashes or flying in improper areas were reported over the past few years, such as the Gatwick airport incident. The airport had to suspend flights in and out for several hours due to the sight of two drones near the runway. While uUAVs are susceptible to all these kinds of risks, tUAVs are physically connected to the GS through the tether. This limited length-tether can add another measure of controllability to the tUAV, which is able to prevent the drone from straying away. In fact, the tether is used in some works specifically to enhance the safety of the tUAV and its resistance to wind and harsh weather conditions. The tether can also be used as an alternative to GPS (global positioning system) to determine the location of the tUAV and ensure a safer landing.

\section*{Use Cases}
The typical use cases for uUAV-mounted BSs are limited to temporary scenarios such as providing coverage for (i) disaster areas with temporarily unavailable infrastructure, or (ii) events like concerts or soccer games. However, given their longer flight times, tUAVs can provide cellular coverage for much more scenarios. They maintain the reliability of terrestrial BSs, in terms of providing uninterrupted service, while bringing the inherent advantages of deployment at high altitude.
\subsection*{Use Case 1: Capacity Enhancement and Traffic Offloading in Dense Urban Areas}
tUAV's main applications and use cases are those that require high endurance and prolonged flight time. For instance, uUAV with a flight time of one hour is eligible for applications like providing cellular coverage in emergency scenarios, or short-term events. However, traffic offloading in urban areas requires a more sustainable UAV operation, which is a perfect fit for the tUAV capabilities. As shown in Fig.~\ref{fig:urban}, GSs can be placed on multiple rooftops in dense urban areas. As noted earlier, the taller the rooftop, the larger the hovering region of the tUAV gets. Having multiple tUAVs with large hovering regions enables changing the constellation of the tUAVs in the sky whenever needed, based on the traffic demands and the user locations. Reaching such flexibility in the spatial distribution of the tUAVs actually reduces the effect of the mobility restrictions induced by the tether, leading to a performance almost similar to that of uUAVs, in terms of mobility and relocation flexibility. 
\subsection*{Use Case 2: Coverage Enhancement in Rural Areas}
Network operators are often not willing to invest in rural and low income areas, due to high costs of network deployment and low potential profits. As briefly discussed earlier, tUAVs require much less time and money for deployment and operation compared to typical terrestrial BSs. As shown in Fig.~\ref{fig:rural}, a tUAV communication system can be used to enhance cellular coverage in rural areas. Due to the nature of rural areas, where there are no much tall buildings, placing the GS on a moving vehicle can be enough to achieve large hovering regions. In addition, since traffic demands are significantly lower than urban counterparts, continuously changing the spatial location of the tUAV will not be necessary. 
\subsection*{Use Case 3: Network Densification}
One of the top benefits of airborne BSs is the quality of A2G channel with mobile users. Due to the higher probability of establishing a LoS A2G channel, the coverage radius of an airborne BS is higher than that of a terrestrial BS. With the stable power supply carried through the tether, enabling long term operation, tUAV can be used for network densification in areas with high traffic demand. Even though tUAV's mobility is restricted compared to uUAV, it still brings the benefits of high altitude deployment. 
\begin{figure}[!t]
    \centering
        \centering
\includegraphics[width=0.8\columnwidth]{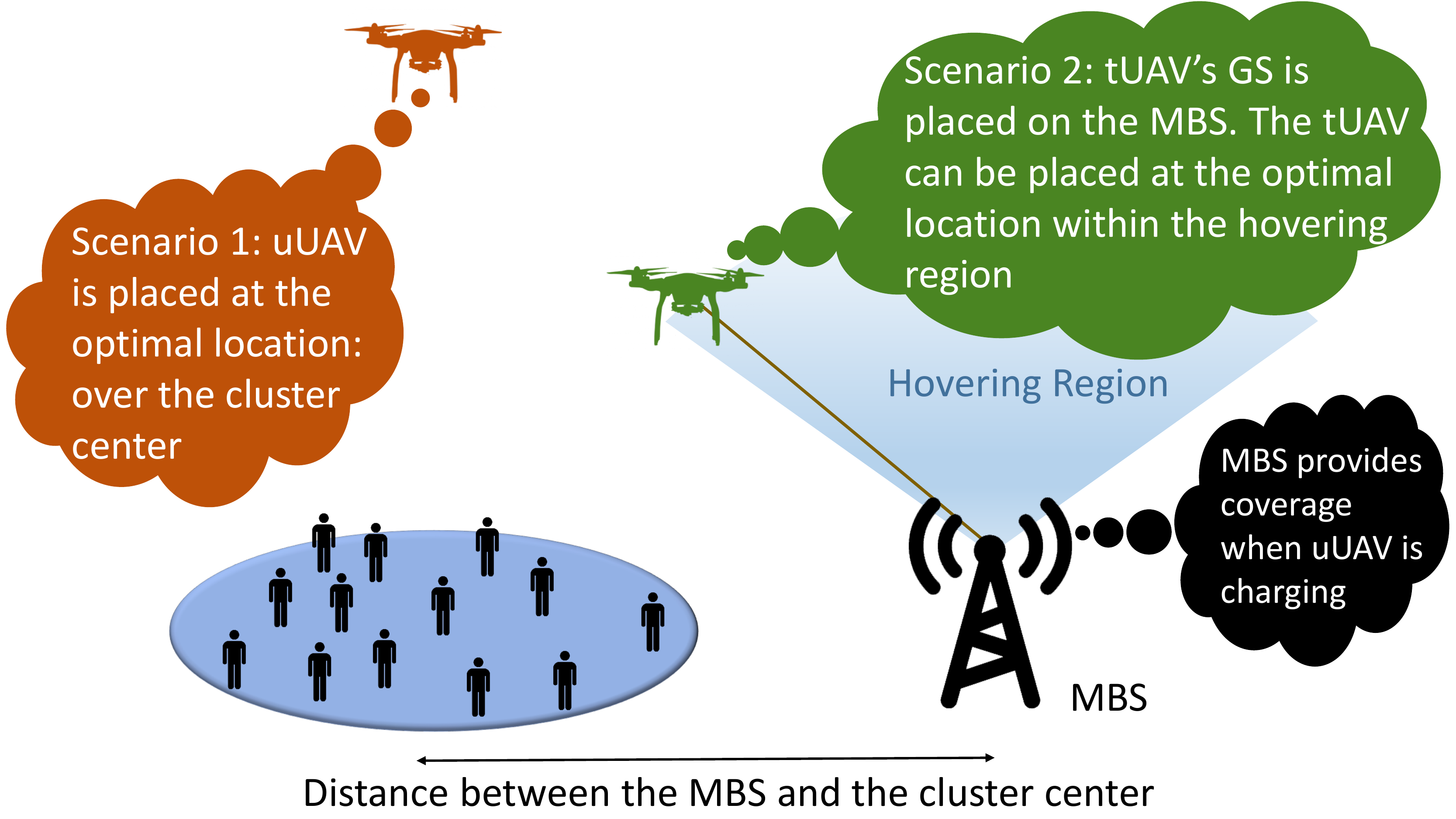}
\caption{System Setup for Fig.~\ref{fig:densification}}
\label{fig:densification:setup}
\end{figure}
  \begin{figure}[!t]
        \centering
\includegraphics[width=0.5\columnwidth]{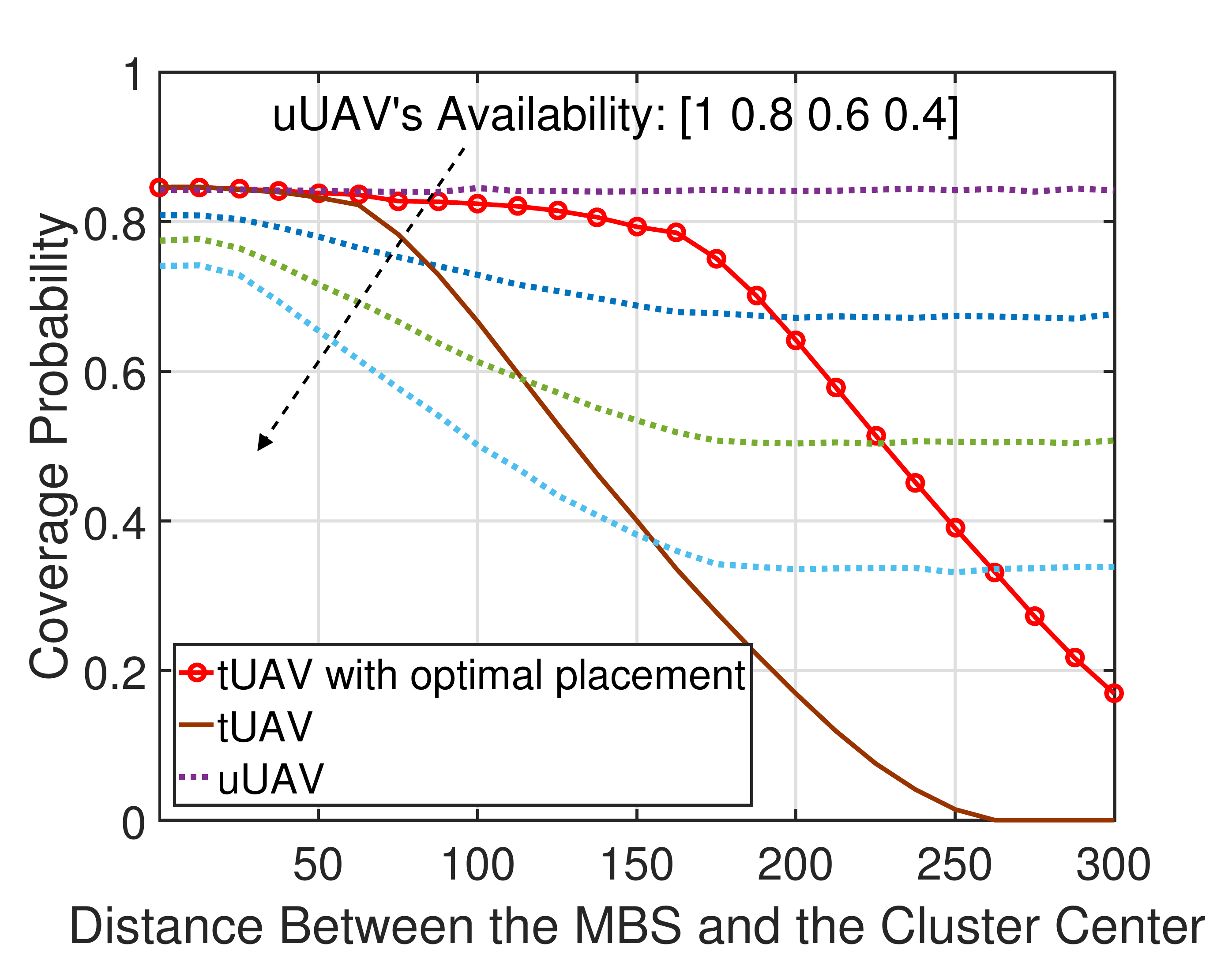}
     \caption{The coverage probability when using either tUAV or uUAV for different values of the distance between the MBS and the cluster center. In this simulation, as shown in Fig.~\ref{fig:densification:setup}, we consider a setup of one circular cluster of users, with the users uniformly distributed inside a disk of radius 100 m. The uUAV's availability defines the fraction of the time at which the uUAV is operating, where there rest of the time it is recharging/swapping its battery. The tUAV is connected to an MBS through a tether with length 120 m~\cite{tds}. We compare 3 scenarios: (i) scenario 1 is when uUAV is used and the main limitation is its availability, (ii) scenario 2 is when tUAV is used and placed at the optimal location within the hovering region, and (iii) scenario 3 is when the tUAV is placed directly above its GS (no optimal placement is considered).}%
      \label{fig:densification}
 \end{figure}

\begin{figure}[!t]
    \centering
        \centering
\includegraphics[width=0.8\columnwidth]{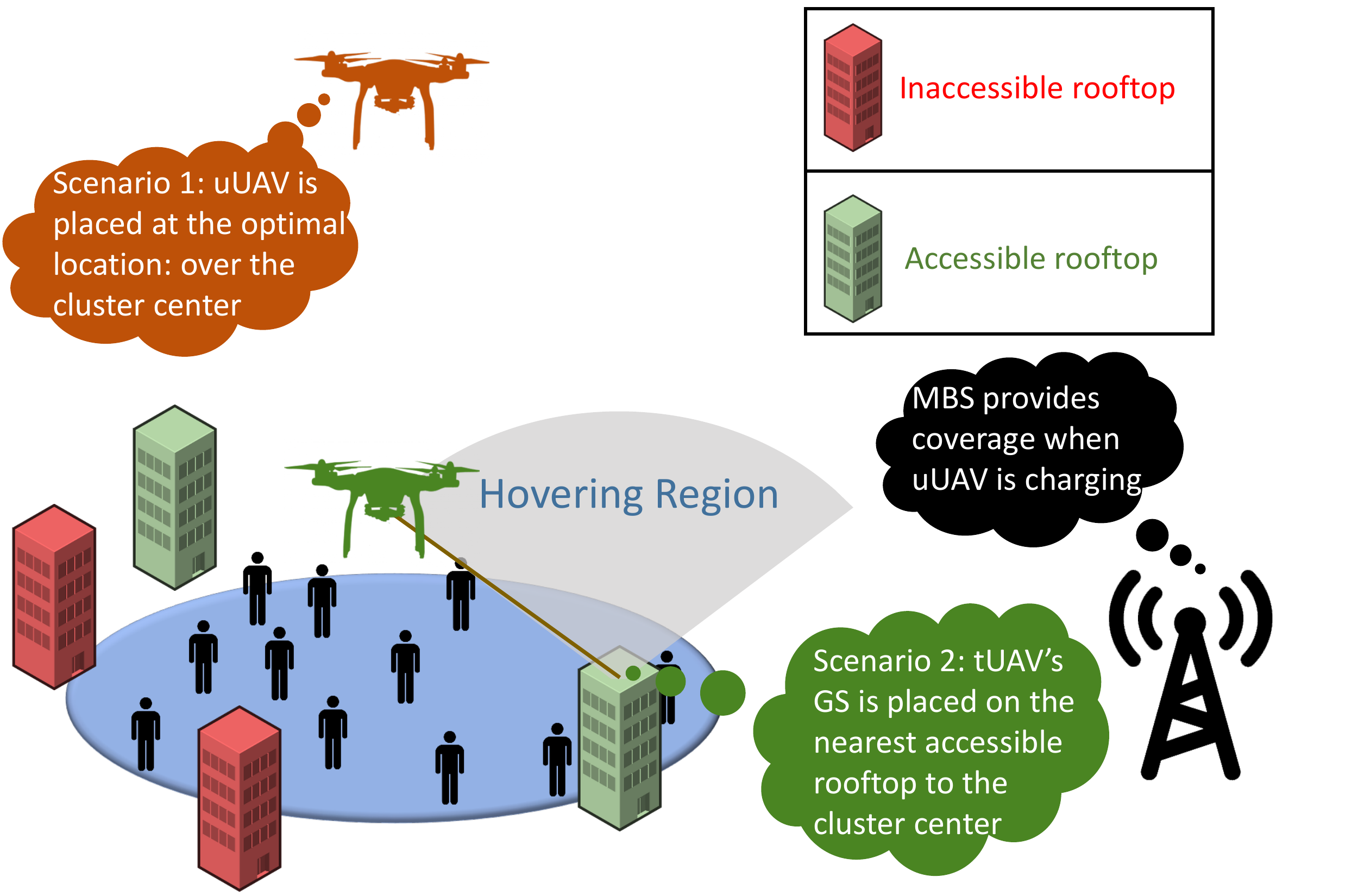}
\caption{System Setup for Fig.~\ref{fig:cluster}}


\label{fig:cluster:setup}
    \end{figure}%
    \begin{figure}[!t]
        \centering
\includegraphics[width=0.5\columnwidth]{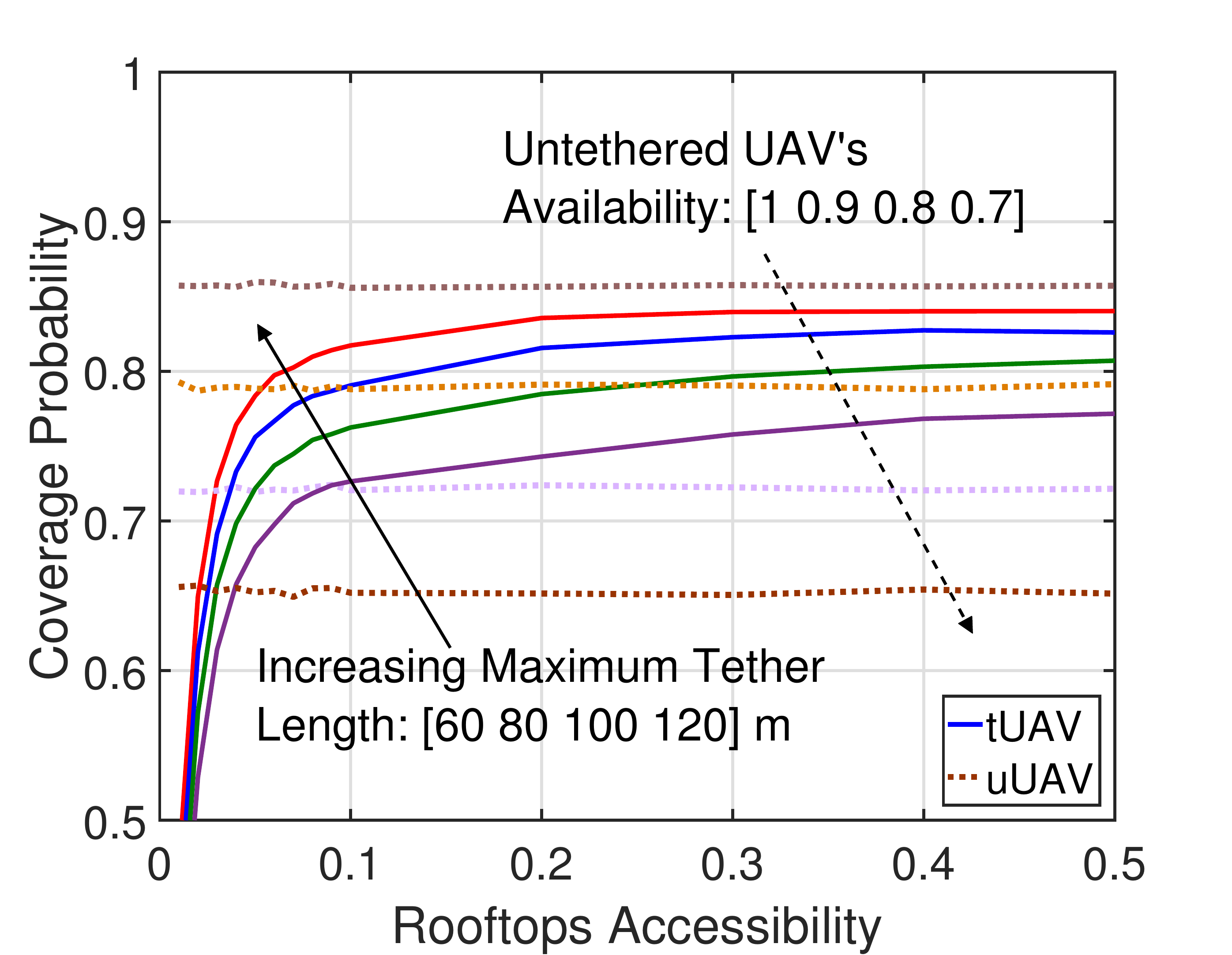}
\caption{The coverage probability for different values of uUAV's availability, maximum tether length, and rooftops accessibility. In this simulation, as shown in Fig.~\ref{fig:cluster:setup}, we study the coverage of uniformly distributed users within a disc of radius 100 m. The MBS is located 160 m away from the disc center. The GS of the tUAV is placed at the nearest accessible rooftop to the disk center. The rooftops accessibility defines the fraction of the buildings where it is permitted to deploy the GS. The density of the buildings is 500 buildings/km$^2$. We compare 2 scenarios: (i) scenario 1 is when uUAV is used and the main limitation is its availability, (ii) scenario 2 is when tUAV is used and the GS is placed at the closest accessible rooftop to the cluster center.}%
\label{fig:cluster}

\end{figure}
\section*{Endurance versus Mobility: Simulations and Discussion}
In this section, we aim to show, with the aid of Monte-Carlo simulations, the trade-off between tUAVs and uUAVs in terms of (i) unconstrained mobility with limited flight time for uUAVs, and (ii) constrained mobility with unlimited flight time for tUAVs. We first consider a system setup composed of a MBS, a cluster of users, and a UAV deployed to serve this cluster of users and offload the MBS. The locations of the users are uniformly distributed inside a cluster of radius 100 m. In case of using an uUAV, we assume that the uUAV hovers at the center of the cluster to maximize coverage. However, due to battery limitations, the uUAV has to leave its aerial location and fly back to a charging station to recharge/replace its battery. During this time, the users are only served by the MBS. Hence, we introduce the uUAV's availability as the fraction of the time where it is actually operating. On the other hand, in case of using a tUAV, we assume that it has unlimited flight time. However, its mobility is limited by a tether connecting it to the MBS with length of 120 m, similar to the tUAV specifications described in~\cite{tds}. Hence, here we are assuming that the tUAV's GS is the MBS. In Fig.~\ref{fig:densification}, we observe that in case of having an uUAV with availability of 1, we will have the best possible coverage. However, as discussed earlier, we don't have the technology to achieve such setup yet. In case of having an availability of 0.8, the tUAV outperforms the uUAV as long as the distance between the MBS and the cluster center is below 193 m. This threshold increases when the availability of the uUAV decreases. 

In Fig.~\ref{fig:densification}, we also compare the performance of the tUAV for two deployment scenarios: (i) the tUAV is hovering exactly above the GS with the tether extended to its maximum value, and (ii) the tUAV is placed at the optimal location within its hovering region that maximizes the coverage probability. The results show the importance of the tUAV's optimal placement. Note that the placement optimization problem of the tUAV is different from the typical 3D placement problems of uUAV discussed in literature. This is mainly due to the restricted mobility of the tUAV, which reduces its reachable 3D locations. Note that this placement problem is different from the scenario of having a maximum allowable altitude for the uUAV. For the latter, the uUAV can hover anywhere as long as it maintains its altitude below a given value, which is not the case for the tUAV. For more details, please refer to~\cite{kishk2019}.

As stated earlier, the GS does not have to be a MBS. It can be the rooftop of any building as long as it has access to a stable energy resource and a reliable connection to the core network. Obviously, these conditions are not always satisfied by any randomly selected building. In addition, not every building satisfying these conditions will grant access to the operator to deploy its GS on their rooftop. Hence, for a given density of buildings, we introduce the rooftops accessibility as the ratio of buildings that satisfy the aforementioned conditions and are willing to grant access to their rooftop. We consider similar setup as in Fig.~\ref{fig:densification} with the GS deployed at the nearest accessible rooftop to the cluster center, instead of deploying it at the MBS. In addition, we fix the distance between the MBS and the cluster center to 160 m. We model the locations of the buildings using a Poisson point process (PPP) with density 500 buildings/Km$^2$, which is the typical density of buildings in urban areas. In Fig.~\ref{fig:cluster}, we compare the performance of uUAV and tUAV for different values of rooftops accessibility. We observe that the minimum required rooftop accessibility for the tUAV to outperform the uUAV decreases as we increases the tether length. For instance, when the availability is 0.9, the required rooftops accessibility decreases from 0.25 to 0.05 as we increase the maximum tether length from 80 m to 120 m. This result shows the influence of the maximum tether length on the system performance. Given that the rooftops accessibility constitutes an important part of the capital expenditure (CAPEX) of the system, these results show that increasing the maximum tether length is actually important for a cost efficient deployment of the tUAVs. 

\section*{Challenges and Design Considerations}
The challenges of uUAV communications have received extensive discussion in literature, specially the issues arising from deployment at such high altitude. Hence, in this section, we focus on the challenges that are related specifically to the proposed setup. In particular, we provide the challenges that exist only in tUAV setups, but not in uUAV ones.

\textbf{Challenge 1}. While airborne communication systems, in general, require new regulatory policies, tUAV systems might need some special considerations. For instance, new safety regulations should be implemented for the areas where tethers are allowed to extend. Safety margins around buildings and above ground have to be kept to avoid (i) any accidents because of tangling or (ii) any malicious attempts to mess with the tether. Given the high importance of the tether in the system, carrying data and providing power to the drone, its safety is vital to the safety of the drone.  These restrictions impose some constraints on the potential deployment locations of the GS and the hovering regions. Hence, the tUAV optimal placement problem should take such safety regulations into consideration.

\textbf{Challenge 2}. As it can be noticed from Fig.~\ref{fig:urban}, it makes more sense to place the GSs on tall rooftops when establishing tUAV systems in dense urban areas, due to the high density of obstructions (tall buildings). On the other hand, as observed from Fig.~\ref{fig:rural}, rural areas are less-obstructed, hence, the altitude of the tUAV does not have to be very high, making it sufficient to place the GS on a moving vehicle. However, when deploying tUAV systems in urban or suburban areas, a trade-off comes to picture. On one hand, placing the GS on a moving vehicle has the advantage of mobility and, hence, the ability to relocate the GS, whenever needed, towards areas with more user density and higher traffic demand. In addition, it is less expensive than rooftops which require monthly/annual rents for building owners. On the other hand, rooftops have the advantage of higher altitude, which adds extra hovering region for the tUAV given the limited tether length. In addition, it keeps the tether away from public access, ensuring safer operation. 

\textbf{Challenge 3}. Unlike typical uUAV placement optimization research work, tUAV placement problem is different. Each tUAV has to be physically connected to the GS on the rooftop through the tether during operation. Hence, the problem is more constrained and needs to be carefully studied. The rooftop selection problem can be solved using different approaches depending on the main objectives of the operator in terms of quality of service (QoS). In addition to cellular coverage-related considerations, cost efficiency should also be taken into consideration during the rooftop selection process. In fact, there is a trade-off between deploying less tUAVs at tall rooftops located in the middle of user hotspots (probably higher rents), and deploying more tUAVs at shorter rooftops. This trade-off between capital expenses (number of tUAVs) and operational expenses (rooftop rents) adds another layer of complexity to the optimal rooftop selection problem.

\textbf{Challenge 4}. Given the location selected for placing the GS, it is important to know exactly how the hovering region looks like. Given the constraints of avoiding tangling upon neighboring buildings, ensuring being far enough from public access, and establishing safety margin above all surrounding buildings for safety, the hovering region of each rooftop is actually unique. For instance, if the rooftop is surrounded by shorter buildings from all sides, it will have a larger hovering region, and hence, more mobility freedom is ensured to the tUAV. The hovering region is a function of the distances to the surrounding buildings and their relative heights. In order to solve the 3D placement optimization problem of a tUAV, an analytical model for the hovering region needs to be derived first. 


\section*{Conclusion}
In this article, we discussed the potential of tUAV for cellular coverage and capacity enhancement. The proposed setup can be thought of as a compromise that aims to replace the current uUAV performance constraints resulting from limited on-board energy with mobility constraints resulting from the tether connection. We showed that tUAV systems have some promising advantages compared to uUAVs, despite the mobility constraints resulting from the tether. We discussed some potential use cases and applications where tUAV-mounted BS will be of great benefit, such as capacity enhancement in urban areas, coverage enhancement in rural areas, and network densification. Finally, we discussed some open challenges and research problems that need to be well-investigated in order to understand better the performance limitations of the proposed setup.
\section*{Acknowledgment}
This work was presented in part in the 43rd WWRF meeting in London.
\bibliographystyle{IEEEtran}
\bibliography{Draft_v0.7.bbl}
\section*{Biographies}
{Mustafa A. Kishk} [S'16, M'18] is a postdoctoral research fellow in the Communication Theory Lab at King Abdullah University of Science and Technology (KAUST). He received his B.Sc. and M.Sc. degree from Cairo University in 2013 and 2015, respectively, and his Ph.D. degree from Virginia Tech in 2018. His current research interests include stochastic geometry, energy harvesting wireless networks, UAV-enabled communication systems, and satellite communications.

{Ahmed Bader} [M'10, SM'13] received the B.S. degree
from the University of Jordan in 2003, the M.S. degree from
OSU in 2006, and the Ph.D. degree
from Telecom ParisTech, in 2013, all in Electrical
Engineering. He holds multiple U.S and EU patents. He is also
a co-founder of Insyab Wireless (www.insyab.com), a Dubai-based
company designing real-time connectivity solutions for
unmanned systems. He received the IEEE ComSoc Young
Professionals Best Innovation Award in 2017.

{Mohamed-Slim Alouini} [S'94, M'98, SM'03, F'09] received his Ph.D. degree in Electrical Engineering (EE) from the California Institute of Technology (Caltech), Pasadena, in 1998. He served as a faculty member at the University of Minnesota, Minneapolis, and then at the Texas A$\&$M University at Qatar, Education City, Doha, before joining KAUST as a Professor of EE in 2009. His current research interests include the modeling, design, and performance analysis of wireless communication systems.
\end{document}